\documentclass[prb, twocolumn, superscriptaddress]{revtex4-2}
\usepackage{amsmath,amssymb,bm}
\usepackage{hyperref}
\usepackage{graphicx}
\usepackage{epstopdf}
\usepackage{latexsym}
\usepackage{braket}
\usepackage{float}
\usepackage[normalem]{ulem}
\usepackage{comment}
\usepackage{mathtools}
\usepackage{array}
\usepackage{tabu}
\usepackage{multirow}
\usepackage[inline]{enumitem}
\usepackage{cleveref}
\usepackage[usenames,dvipsnames,table,xcdraw]{xcolor}
\usepackage[normalem]{ulem}
\def\BibTeX{{\rm B\kern-.05em{\sc i\kern-.025em b}\kern-.08em
    T\kern-.1667em\lower.7ex\hbox{E}\kern-.125emX}} 

\begin{document}
\title{Statistical Interaction Driven Thermoelectricity and Violation of Wiedemann–Franz Law}
\author{Sampurna Karmakar}
\affiliation{Indian Institute of Science Education and Research Kolkata, Mohanpur, Nadia - 741246, West Bengal, India}
\author{Amulya Ratnakar}
\affiliation{Aix Marseille Univ, Université de Toulon, CNRS, CPT, Marseille, France}
\author{Sourin Das}
\affiliation{Indian Institute of Science Education and Research Kolkata, Mohanpur, Nadia - 741246, West Bengal, India}
\begin{abstract}
Quantum transport anomalies in systems obeying Haldane–Wu fractional exclusion statistics, characterized by the statistical interactions parameter $g$ are investigated. We identify particle–hole symmetry breaking of the Haldane–Wu distribution function via its deviations of the maximum entropy ($\mathcal{S}_{g}^{max}$), evaluated at the chemical potential, from the value ${k_B} \ln 2$ (a value that holds only at the free fermion limit, $g=1$). A duality relation, $g\,\mathcal{S}_{g}^{max}=\mathcal{S}_{1/g}^{max}$,  quantifying the degree of violation is obtained.  This symmetry breaking manifests in transport phenomena as: significant violations of the Wiedemann–Franz law arising for $g>1$ (but remain absent for $g\leq 1$) across a broad temperature range. Moreover, the thermoelectric figure of merit $ZT$ is substantially enhanced for $g>1$ and suppressed for $g<1$, indicating new routes to optimize energy conversion. These results deepen the understanding of the interplay between equilibrium statistics and transport, suggesting avenues for engineering advanced thermoelectric materials.
\end{abstract}
\maketitle
Fermions and bosons are fundamental particles that obey distinct quantum statistics: Fermi-Dirac for fermions and Bose-Einstein for bosons. This dichotomy can be extended by introducing frameworks that allow for a continuous interpolation between the two, enriching our understanding of quantum statistics. This interpolation can be realized in two main ways: \\
{\it{(i)}} by defining abelian braid statistics in two-dimensional systems, where the many-body wavefunction acquires a non-trivial phase (between $0$ and $\pm 2\pi$) upon particles being adiabatically taken around each other. This leads to the concept of anyons~\cite{Anyon1,Anyon2,Anyon3,Anyon4},  leading to a contentious interpolation between the two extremes (zero being a boson and $2\pi$ being a fermion),\\
{\it{(ii)}} through a single particle state-counting approach developed by Haldane~\cite{Haldane_exclusion}, for a system with fixed size and boundary conditions. It was noted that the number of available single-particle states for the $(N+1)^{th}$ particle depends on both $N$ and the underlying particle statistics leading to an identification of a statistical interaction parameter $g$. By allowing for fractional occupancy of single particle states governed by a generalized exclusion principle, this framework leads to Haldane exclusion statistics (HES), providing a statistical description beyond the conventional binary classification. Unlike the case with {\it{(i)}}, this is not restricted to two-dimensional systems.

It is well-established that anyons in strong magnetic fields, when projected onto the lowest Landau level, follow HES~\cite{PhysRevLett.72.600}, hence providing a unification point for the two approaches described above. Later, it was shown that an incompressible anyon liquid in the presence of a Hall response exhibits HES, and its implications for edge states were also discussed~\cite{PhysRevB.92.235151}. Also, the ideal gas of HES particles and its interacting counterpart in one dimension can be bosonized and exhibit the universal properties dictated by Luttinger liquid fixed point at low temperatures~\cite{murthy1,PhysRevLett.75.890,WU2001551}. These intimate connections of HES to the anyonic quantum Hall state and Luttinger liquids (whose chiral counterpart describes the edge states of the quantum Hall state) have motivated studies~\cite{Fractional_thermal,Martin} in the past pertaining to transport properties of these particles in a Hall bar type geometry and is the focus of this letter.

\begin{figure}[t]
    \centering
    \includegraphics[width=0.9\linewidth]{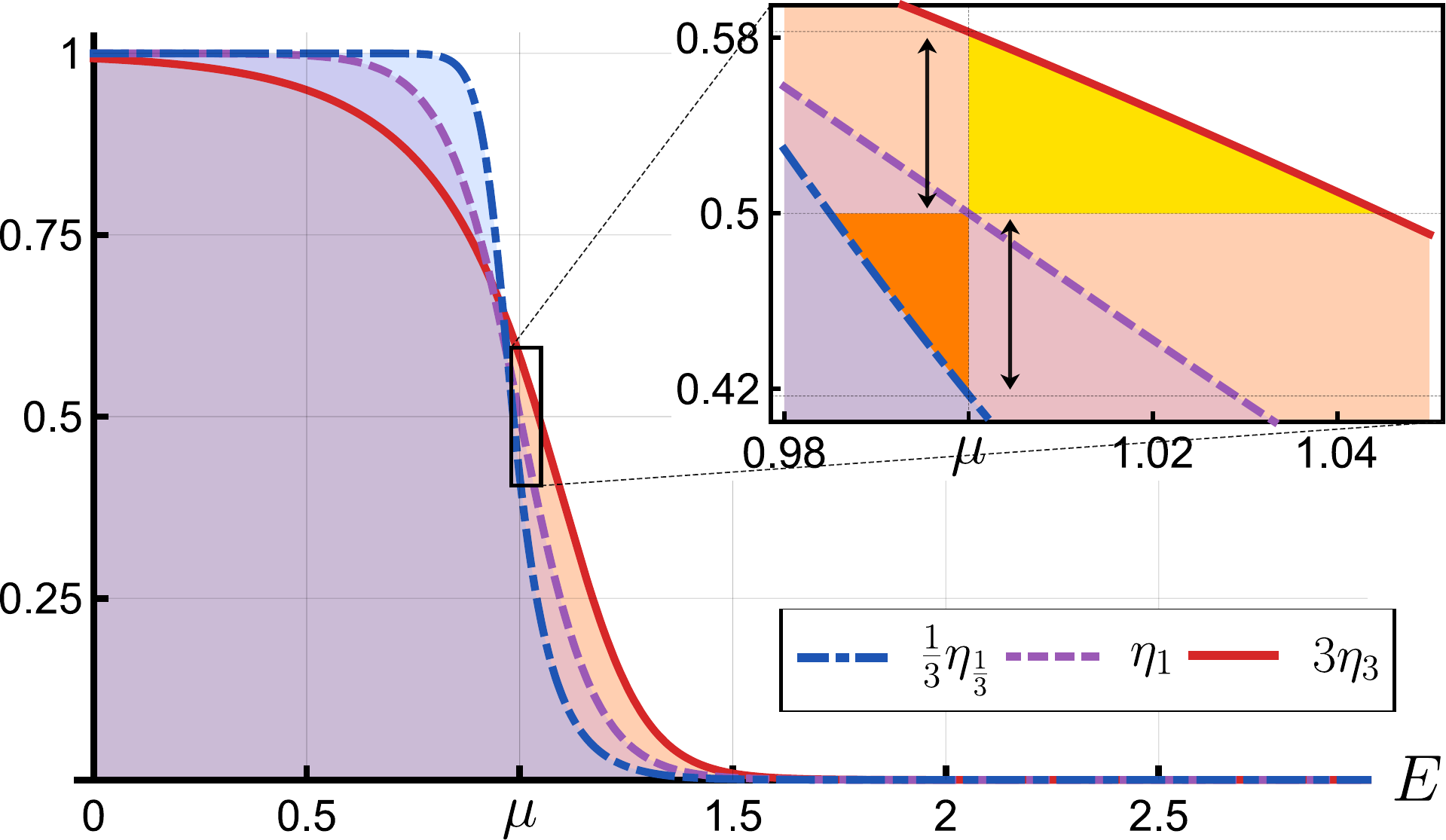}
    \caption{Occupation functions at finite temperature ($T=1K$), scaled by a factor of $g$ for visual comparison, obtained from HES~\cite{Wu}, are shown for three different statistical parameter values:  $g=1/3$ (blue dash-dotted line), $g=1$ (violet dashed line) and $g=3$ (red solid line), with the corresponding areas under the curves shaded. In the inset, the regions indicating particle–hole symmetry breaking for $g=1/3$ and $g=3$ are marked in orange and yellow, respectively. The up down arrows highlight equal and opposite deviations from the particle–hole symmetric point for $g=1$ at $E=\mu$, as illustrated by eq.~(\ref{eq:deviation}).}
   \label{fig:Distribution_function}
\end{figure}

\textit{Haldane exclusion statistics and particle-hole symmetry breaking}:- The concept of fractional statistics was introduced by Haldane~\cite{Haldane_exclusion}, who proposed that the change in the number of available single-particle states ($\Delta d$) in a system with fixed size and boundary conditions is related to the change in particle number ($\Delta N$) as: $\Delta d=-g\,\Delta N$. The statistical parameter $g$ is assumed to be rational, with $g = 0$ and $g = 1$ corresponding to bosonic and fermionic statistics, respectively. This generalized Pauli’s principle leads to an occupation function obtained by Wu~\cite{Wu} for an ideal gas of particles obeying fractional statistics, given by
\begin{align}
    \eta_g(E;\mu,T)=1/\left[\mathcal{W}(x,g)+g\right],
    \label{Eq:distribution}
\end{align}
and $\mathcal{W}(x,g)$ satisfies: $  \mathcal{W}(x,g)^g\,[1+\mathcal{W}(x,g)]^{1-g}=e^{x}$, where $x=(E-\mu)/(k_\text{B} T)$, $E$ is energy of a particle, $T$ is temperature, $\mu$ is the chemical potential. $\eta_g (E; \mu, T)$ exhibits dependence in the form $\eta_g ((E- \mu)/T)$, and the duality relations between $g$ and $1/g$ obtained by Nayak and Wilczek reads as ~\cite{Chetan}
\begin{align}
    1-g\,\eta_g\left((E- \mu)/T\right)=(1/g)\,\eta_{1/g} ((-1/g)(E- \mu)/T).\label{eq:duality}
\end{align}
For $g\neq 0$, at zero temperature, the occupation function of single-particle states exhibits a step-like behavior, clearly indicating the presence of a well-defined Fermi level. Specifically, it takes the value $1/g$ for energy levels below the chemical potential ($ E < \mu$) and drops to zero for $ E > \mu$. The introduction of an additional particle into the system results in the occupation of $g$ single-particle states.
\begin{figure}[t]
    \centering
    \includegraphics[width=1\linewidth]{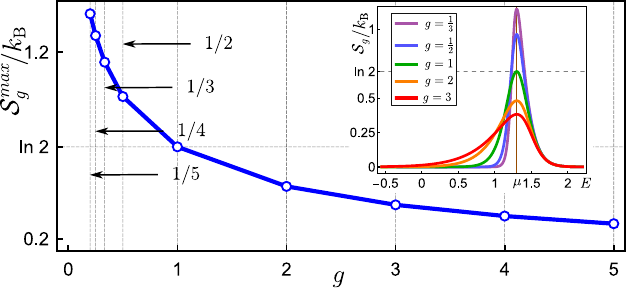}
    \caption{Maximum entropy for different values of $g$ with the entropy profile at $T=1K$ shown in the inset.}
    \label{fig:entropy}
\end{figure}
The entropy associated with such particles is~\cite{Wu,ISAKOV1}
\begin{align}
    \mathcal{S}_g=-k_{\text{B}}\big[\eta_g&\ln\eta_g+(1-g\,\eta_g)\ln(1-g\,\eta_g)\nonumber\\
    &-(1+(1-g)\,\eta_g)\ln(1+(1-g)\,\eta_g)\big],
\end{align}
which reaches its maximum value at $E=\mu$ (see inset of fig.~\ref{fig:entropy} ). Note that this value is independent of temperature, as expected. A corresponding duality relation (following eq.~(\ref{eq:duality})) is obtained as
\begin{align}
   \mathcal{S}_{g}^{max}=k_{\text{B}}\ln \mathcal{W}(0,g)/(g-1) \text{ and } g\,\mathcal{S}_{g}^{max}=\mathcal{S}_{1/g}^{max}.
\end{align}
Note that, for the limiting case with $g=1$, we have $\mathcal{W}(0,1)=1$, which leads to
\begin{align}
   \mathcal{S}_{g=1}^{max}/k_{\text{B}}=\ln \mathcal{W}(0,g)/(g-1)=\ln(1+\mathcal{W}(0,g))/g=\ln 2.\nonumber
\end{align}
This value has a clear physical interpretation in terms of counting of available states : at $E=\mu$, the occupation probability is $\eta_1|_{E=\mu}=1/2$ (see fig.~\ref{fig:Distribution_function}), implying that a single-particle state for free fermions $(g = 1)$ can be either occupied or empty with equal probability at $E=\mu$, resulting in an entropy of $k_{\text{B}}\ln 2$. 

This behavior is identified as particle–hole symmetry of the occupancy function. Any deviation from this value of entropy indicates a change in the effective degeneracy (number of available states) at the Fermi level, and hence a departure from particle–hole symmetry. For $g\neq 1$, the maximum entropy deviates from $k_{\text{B}}\ln 2$ (see fig.~\ref{fig:entropy}), signaling the absence of particle–hole symmetry. This asymmetry is further illustrated in fig.~\ref{fig:Distribution_function}. The dash-dotted blue line for $g=1/3$ is clearly showing a suppression, and the solid red line for $g=3$ is showing an equal enhancement for its occupancy scaled by $g$ at $E=\mu$ with respect to the $g=1$ case. The duality relation in eq.~(\ref{eq:duality}), when evaluated at $E=\mu$, clearly highlights the structure of the asymmetry with respect to $g=1$, and yields the relation
\begin{align}
    g\,\eta_g\big|_{E=\mu}-\eta_1\big|_{E=\mu}=\eta_1\big|_{E=\mu}-(1/g)\,\eta_{1/g}\big|_{E=\mu}.\label{eq:deviation}
\end{align}
This observation forms the cornerstone of our study. The breaking of particle-hole symmetry about the Fermi level directly suggests the possibility for thermoelectric generation, which is also expected to exhibit sign reversal under the duality transformation ($g \leftrightarrow 1/g$). This conclusion draws motivation from Ref.~[\onlinecite{Viola}] in the context of the fractional quantum Hall system, where the many-body degeneracy resulted in breaking of particle-hole symmetry, and leads to a finite thermoelectric current, which was experimentally verified by Gurman et al.~\cite{Gurman2012}. 
In what follows, we demonstrate how this particle-hole symmetry breaking leads to a large thermoelectric effect and violation of the Wiedemann-Franz law.

\textit{Transport coefficient}:- We consider two infinitely extended counter-propagating chiral one-dimensional modes, characterized by particles of statistical parameter $g$, in the presence of a localized scattering potential at the centre similar to that of a quantum point contact (QPC) in a Hall bar geometry. The two incoming modes are assumed to be in equilibrium with reservoirs maintained at chemical potentials (temperatures), $\mu_1 (T_1)$ and $\mu_2 (T_2)$, respectively. The chemical potentials are related to the voltages $V_1$ and $V_2$ via $\mu_i=qV_i$, where $q$ denotes the charge of the particles. We assume $\mu_{i}=\mu+\delta \mu_i$ and $T_{i}=T+\delta T_i$ for $i=1,2$, where $\mu$($T$) denotes the average chemical potential (temperature) of the system and $\delta\mu_i, k_\text{B}\,\delta T_i$ are assumed to be very small compared to $k_\text{B}\, T$ and $\mu$. We consider the linear response limit, where the applied chemical potential bias $\mu_{1}-\mu_{2}=\delta \mu_{1}-\delta\mu_{2}=\Delta \mu=q\Delta V  \longrightarrow 0$ and temperature difference $T_1-T_2=\delta T_1-\delta T_2=\Delta T\longrightarrow 0$. The QPC is considered to be partially transmitting, where particles incident along the incoming modes are transmitted or reflected to the outgoing modes with energy-dependent probabilities $t(E) $ and $r(E)$, respectively, satisfying $t(E)+r(E) =1$ to ensure probability conservation. It has been demonstrated in Ref.~[\onlinecite{Krive}] that the transport coefficients for particles obeying exclusion statistics follow the same general expressions as those for particles with conventional statistics. Hence, we can write (see appendix~\ref{sec:Linear_transport_coefficients})
\begin{align}
    \begin{pmatrix}
        I_{net}\\J_{net}\end{pmatrix}=\begin{pmatrix}          
       q^2 \mathcal{L}_0 &q \mathcal{L}_1\\
        q\mathcal{L}_1 & \mathcal{L}_2
    \end{pmatrix}\begin{pmatrix}
        \Delta\mu/q\\ \Delta T /T\end{pmatrix},\label{Eq:Onsager_matrix1}
\end{align}
where $I_{net}$ and $J_{net}$ are the net charge and heat current flowing through the system from left to right. The Onsager coefficients $\mathcal{L}_\alpha (\alpha=0,1,2)$ are defined as
\begin{align}
    \mathcal{L}_\alpha=\frac{1}{h}\int_{E(0)}^\infty dE \,t(E)(E-\mu)^\alpha\big(-\eta_g'(E;\mu,T)\big),
\end{align}
where $'$ represents the first-order derivative with respect to energy. 
Eq.~(\ref{Eq:Onsager_matrix1}) satisfy the  Onsager reciprocal relations in accordance with fundamental thermodynamic principles.
\begin{figure}[t]
    \centering
    \includegraphics[width=0.95\linewidth]{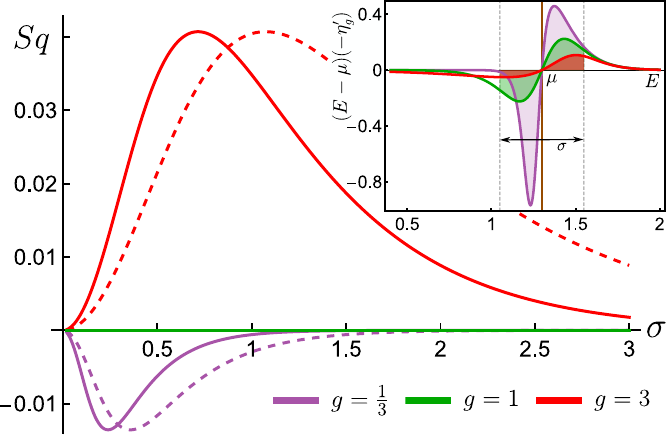}
    \caption{The Seebeck coefficient multiplied by the particle charge, $Sq$, is plotted as a function of the width $\sigma$ of the energy-dependent transmission function $t(E)$, defined in eq.~(\ref{Eq:t1(E)}). Results are shown for three values of the statistical parameter: $g = 1/3$ (purple),  $g = 1$ (green), and $g =3$ (red). Solid lines correspond to temperature $T=1K$, and dashed lines to $T=1.5K$. In the inset, the functions $(E-\mu)(- \eta'_g)$ are plotted at $T=1K$ for the same $g$ values, with the chemical potential set to $\mu=1.3$ meV (indicated by the brown vertical line). The width of the transmission window $\sigma=0.5$ meV is shown by vertical dashed lines, centered at $\mu$. The product functions $t(E)(E-\mu)(- \eta'_g)$, which we need to determine the Seebeck coefficient, are represented by the shaded areas for each value of $g$.}
    \label{fig:thermoe}
\end{figure}

Now we focus on the case where the transmission function is symmetric with respect to the system’s chemical potential, $\mu$ so that the only source of particle-hole symmetry breaking is the distribution function. This scenario, exemplified by a Lorentzian transmission profile in the Breit–Wigner form~\cite{Supriyo,Alhassid,Nakpathomkun,Kennes_2013,Whitney_review}, is well motivated by transport experiments through a resonant level. In this case the Onsager coefficients confirm to neat duality relations given by 
\begin{align}
    \mathcal{L}_{\alpha,g}(T)&=\frac{(-1)^{\alpha}}{g^2} \mathcal{L}_{\alpha,1/g}(gT),\label{eq:L_dual}
\end{align}
 leading to the duality relations for the longitudinal electrical conductance ($G$), longitudinal thermal conductance ($\kappa$) and Seebeck coefficient ($S$) (see appendix~\ref{sec:Linear_transport_coefficients})
\begin{align}
    G_g(T)&=(1/g^2)G_{1/g}(gT),\label{eq:G_dual}\\
    \kappa_g(T)&=(1/g)\kappa_{1/g}(gT),\label{eq:kappa_dual}\\
     S_g(T)&=-g\,S_{1/g}(gT).\label{eq:seeb_dual}
\end{align}
It is clear from the sign change of the $S_g(T)$ under $g \longleftrightarrow 1/g$ that there is a switching between particle dominated transport and hole dominated transport under duality transformation. 

\begin{figure}[t]
    \centering
    \includegraphics[width=0.98\linewidth]{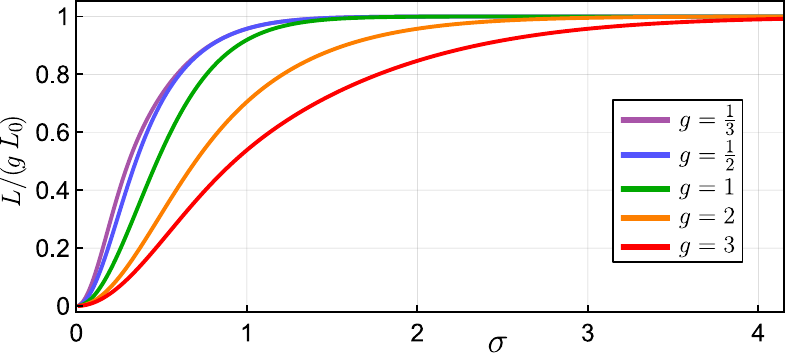}
    \caption{The Lorenz number, normalized by  $g\Bar{L}_0$, is plotted for various values of $g$ using the energy-dependent transmission function defined in eq.~(\ref{Eq:t1(E)}), by varying the width $\sigma=$. The system's chemical potential is at $\mu=1.3meV$  and temperature is $1K$.}
    \label{fig:Lorenz_no}
\end{figure}
\textit{Thermoelectric response}:- We examine the Seebeck coefficient, which characterizes the thermoelectric response, a sensitive probe of asymmetries in the energy dependence of transport. It is defined as $S=-(\Delta V/\Delta T)\big|_{I_{net}=0}=(1/qT)(\mathcal{L}_1/\mathcal{L}_0)$. Since $\mathcal{L}_1$ vanishes for an energy-independent transmission function, any finite thermoelectric signal in such a case must arise from an interplay between the transmission profile and asymmetries in the occupation function. To examine the implications of this asymmetry, we next consider a simple energy-dependent transmission function~\cite{boxcar} which is symmetric about $\mu$:
\begin{align}
   t(E) =
\begin{cases}
1 & \text{for } |E-\mu| \leq \frac{\sigma}{2}, \\
0 & \text{otherwise},
\end{cases}\label{Eq:t1(E)}
\end{align}
\begin{figure*}[t]
    \centering
    \includegraphics[width=0.99\linewidth]{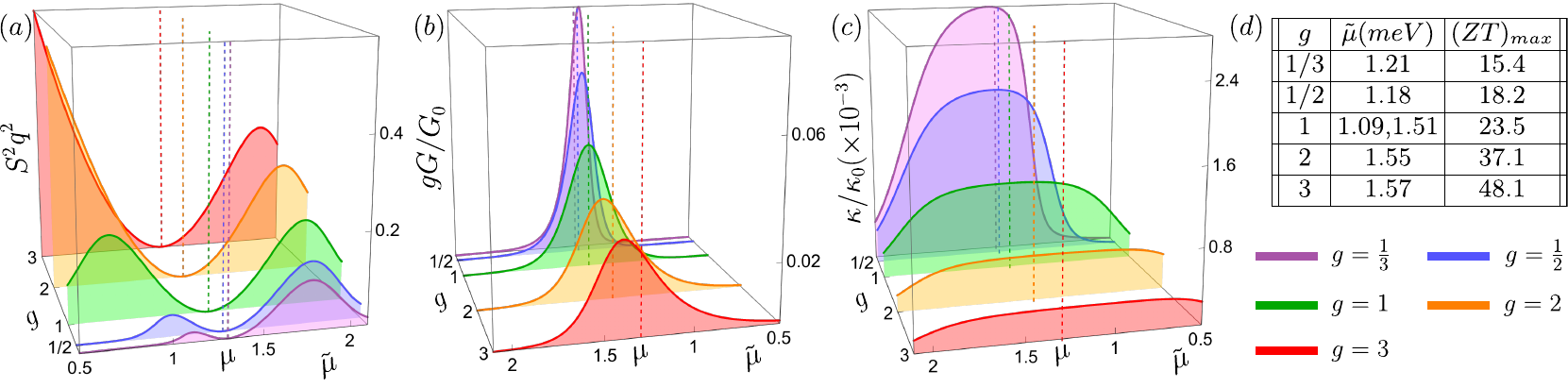}
    \caption{(a), (b), and (c) show the variation of $S^2q^2$, the normalized electrical conductance $gG/G_0$, and the normalized thermal conductance $\kappa/\kappa_0$, respectively, as functions of $\tilde{\mu}$ (resonant level of $t_{qd}(E)$). All plots are evaluated at temperature $T=1K$, with the system's chemical potential fixed at $\mu=1.3$ meV and the parameter $\Gamma=0.01$ meV. Five values of  $g$ are considered in each plot. Panel (d) presents a table listing the maximum values of the thermoelectric figure of merit \textit{ZT} for each $g$, obtained by varying $\tilde{\mu}$.}
    \label{fig:ZT}
\end{figure*}
and we examine how the Seebeck response varies as a function of the transmission width $\sigma$ for $g = 1/3,1$ and $g = 3$ as shown in fig.~\ref{fig:Distribution_function}. This enables a neat investigation of how asymmetry in the occupation function around the chemical potential influences the thermoelectric transport. And the choice of the three $g$ values probes the duality relation. As shown by the green line in the inset of fig.~\ref{fig:thermoe}, the function $(E-\mu)(-\eta'_g)$ is antisymmetric about $\mu$ for $g=1$. Since the transmission function is symmetric about $\mu$, the product $t(E)(E-\mu)(-\eta_g')$ remains anti-symmetric, leading to a vanishing Seebeck coefficient, as reflected by the green line in the main panel of fig.~\ref{fig:thermoe}. In contrast, for $g=1/3$ and $3$, the function $(E-\mu)(-\eta'_g)$ becomes asymmetric (and not anti-symmetric) around the chemical potential, yielding a nonzero Seebeck coefficient, though with opposite signs for $g>1$ and $g<1$  owing to duality. 

\textit{Lorenz number}:- According to the Wiedemann–Franz law, the Lorenz number, defined as $ L=(\kappa/G T)$, attains a universal value in Fermi liquids: $\Bar{L}_0=(\pi^2k_\text{B}^2)/(3q^2)$. In the simplest case where $t(E)$ and $r(E)$ are energy independent, the transport integrals reduce to: $\mathcal{L}_0=(1/ hg)\,t,\, \mathcal{L}_1=0,\, \mathcal{L}_2=(\pi^2k_B^2T^2/3h)\,t$. Here, we have assumed the lower energy bound $E(0)=-\infty$, which is justified as long as it lies well below the thermal excitation window. Under these conditions,  the Lorenz number in the ballistic regime scales with the statistical parameter $g$ as $L=g\Bar{L}_0$, indicating that the Wiedemann–Franz law is obeyed up to a scale factor. This result mirrors the expression obtained in Refs.~[\onlinecite{kane1996thermal},\onlinecite{Etyan_Sourin_2009}] for fractional quantum Hall edge states, where the filling factor $\nu$ can be identified with $1/g$. 

Although this scaling $L=g\Bar{L}_0$ puts the Lorenz number on equal footing under the exchange of $g\leftrightarrow 1/g$ in the ballistic limit, the underlying occupation functions are not. In particular, the asymmetries in the energy derivative of the occupation function, e.g., as illustrated by the plot of $(E-\mu)\big(-\eta_g'(E;\mu,T)\big)$ in the inset of fig.~\ref{fig:thermoe}, differ significantly between $g$ and $1/g$. From eq.~(\ref{eq:G_dual}) and (\ref{eq:kappa_dual}), we can derive a simple expression for the duality relation of the Lorenz number as
\begin{align}
    \frac{L_g(T)}{g\,\Bar{L}_0}=\frac{L_{1/g}(gT)}{(1/g)\Bar{L}_0},
    \label{Eq:L}
\end{align}
with the transmission function being symmetric with respect to the system’s chemical potential. For the transmission function defined in eq.~(\ref{Eq:t1(E)}), we now analyze the deviation of the rescaled Lorenz number $L/g\Bar{L}_0$ from its ballistic value of unity for various values of $g$. As shown in fig.~\ref{fig:Lorenz_no}, asymmetries in the occupation function lead to very different rate at which $L/g\Bar{L}_0$ approaches unity as a function of $\sigma$ for different values of $g$. The $g<1$ case approaches unity much faster than the $g>1$, hence leading to possibility for violations of the Wiedemann–Franz law for $g > 1$ while such violation remaining absent for $g < 1$ across a finite temperature range for an appropriately chosen value of $\sigma$. 

\textit{Figure of merit}:- Next, we turn to another important quantity, the thermoelectric figure of merit, denoted as \textit{ZT}, a dimensionless measure of how effectively a thermoelectric material converts thermal energy into electrical energy. It is defined as:  \textit{ZT}$=S^2GT/\kappa=S^2/L$. We evaluate \textit{ZT} using the transmission function for a scattering region modeled by a quantum dot characterized by a resonance at $\tilde{\mu}$ and width $\Gamma$~\cite{Whitney}: $ t_{qd}(E)=\Gamma^2/[\Gamma^2+4(E-\tilde{\mu})^2]$. By tuning $\tilde{\mu}$ via an external gate voltage, we search for the maximum of the figure of merit for different values of $g$. It is important to note that the quantities involved are interdependent, making the optimization of \textit{ZT} not immediately apparent. Fig.~\ref{fig:ZT} (a), (b), and (c) show the variation of the three constituent quantities---$S^2q^2$, $gG/G_0$ and $\kappa/\kappa_0$, as functions of $\tilde{\mu}$. The factor $g$ is included in the conductance term solely for visual scaling, and the reference values for electrical conductance is $G_0=q^2/h$ and for thermal conductance is $\kappa_0=\pi^2k_B^2T/3h$. Each of these quantities plays a role in determining the thermoelectric efficiency of energy conversion encoded in \textit{ZT}. The corresponding maximum values of \textit{ZT} within this parameter regime for different $g$ values are summarized in the table in fig.~\ref{fig:ZT}. These results reveal a substantial enhancement of the thermoelectric figure of merit, highlighting the potential for efficient thermoelectric performance.
\begin{acknowledgments}
S.K. acknowledges support from the Prime Minister's Research Fellowship (PMRF) scheme of the Ministry of Education, Government of India (PMRF ID: 0501977). A.R. is supported by “ANYHALL” (Grant ANR No. ANR-21-CE30-0064-03). A.R. also received support from the French government under the France 2030 investment plan, as part of the Initiative d’Excellence d’Aix-Marseille Université—A*MIDEX.  S.D. would like to acknowledge warm hospitality recieved from IISER Pune during the finishing stages of this work.
\end{acknowledgments}
\bibliography{citations}

\begin{thebibliography}{29}%
\makeatletter
\providecommand \@ifxundefined [1]{%
 \@ifx{#1\undefined}
}%
\providecommand \@ifnum [1]{%
 \ifnum #1\expandafter \@firstoftwo
 \else \expandafter \@secondoftwo
 \fi
}%
\providecommand \@ifx [1]{%
 \ifx #1\expandafter \@firstoftwo
 \else \expandafter \@secondoftwo
 \fi
}%
\providecommand \natexlab [1]{#1}%
\providecommand \enquote  [1]{``#1''}%
\providecommand \bibnamefont  [1]{#1}%
\providecommand \bibfnamefont [1]{#1}%
\providecommand \citenamefont [1]{#1}%
\providecommand \href@noop [0]{\@secondoftwo}%
\providecommand \href [0]{\begingroup \@sanitize@url \@href}%
\providecommand \@href[1]{\@@startlink{#1}\@@href}%
\providecommand \@@href[1]{\endgroup#1\@@endlink}%
\providecommand \@sanitize@url [0]{\catcode `\\12\catcode `\$12\catcode
  `\&12\catcode `\#12\catcode `\^12\catcode `\_12\catcode `\%12\relax}%
\providecommand \@@startlink[1]{}%
\providecommand \@@endlink[0]{}%
\providecommand \url  [0]{\begingroup\@sanitize@url \@url }%
\providecommand \@url [1]{\endgroup\@href {#1}{\urlprefix }}%
\providecommand \urlprefix  [0]{URL }%
\providecommand \Eprint [0]{\href }%
\providecommand \doibase [0]{https://doi.org/}%
\providecommand \selectlanguage [0]{\@gobble}%
\providecommand \bibinfo  [0]{\@secondoftwo}%
\providecommand \bibfield  [0]{\@secondoftwo}%
\providecommand \translation [1]{[#1]}%
\providecommand \BibitemOpen [0]{}%
\providecommand \bibitemStop [0]{}%
\providecommand \bibitemNoStop [0]{.\EOS\space}%
\providecommand \EOS [0]{\spacefactor3000\relax}%
\providecommand \BibitemShut  [1]{\csname bibitem#1\endcsname}%
\let\auto@bib@innerbib\@empty
\bibitem [{\citenamefont {Leinaas}\ and\ \citenamefont
  {Myrheim}(1977)}]{Anyon1}%
  \BibitemOpen
  \bibfield  {author} {\bibinfo {author} {\bibfnamefont {J.~M.}\ \bibnamefont
  {Leinaas}}\ and\ \bibinfo {author} {\bibfnamefont {J.}~\bibnamefont
  {Myrheim}},\ }\bibfield  {title} {\bibinfo {title} {On the theory of
  identical particles},\ }\href {https://doi.org/10.1007/BF02727953} {\bibfield
   {journal} {\bibinfo  {journal} {Il Nuovo Cimento B (1971-1996)}\ }\textbf
  {\bibinfo {volume} {37}},\ \bibinfo {pages} {1} (\bibinfo {year}
  {1977})}\BibitemShut {NoStop}%
\bibitem [{\citenamefont {Wilczek}(1982)}]{Anyon2}%
  \BibitemOpen
  \bibfield  {author} {\bibinfo {author} {\bibfnamefont {F.}~\bibnamefont
  {Wilczek}},\ }\bibfield  {title} {\bibinfo {title} {Quantum mechanics of
  fractional-spin particles},\ }\href
  {https://doi.org/10.1103/PhysRevLett.49.957} {\bibfield  {journal} {\bibinfo
  {journal} {Phys. Rev. Lett.}\ }\textbf {\bibinfo {volume} {49}},\ \bibinfo
  {pages} {957} (\bibinfo {year} {1982})}\BibitemShut {NoStop}%
\bibitem [{\citenamefont {Wu}(1984)}]{Anyon3}%
  \BibitemOpen
  \bibfield  {author} {\bibinfo {author} {\bibfnamefont {Y.-S.}\ \bibnamefont
  {Wu}},\ }\bibfield  {title} {\bibinfo {title} {General theory for quantum
  statistics in two dimensions},\ }\href
  {https://doi.org/10.1103/PhysRevLett.52.2103} {\bibfield  {journal} {\bibinfo
   {journal} {Phys. Rev. Lett.}\ }\textbf {\bibinfo {volume} {52}},\ \bibinfo
  {pages} {2103} (\bibinfo {year} {1984})}\BibitemShut {NoStop}%
\bibitem [{\citenamefont {Marchetti}(2010)}]{Anyon4}%
  \BibitemOpen
  \bibfield  {author} {\bibinfo {author} {\bibfnamefont {P.~A.}\ \bibnamefont
  {Marchetti}},\ }\bibfield  {title} {\bibinfo {title} {Spin-statistics
  transmutation in quantum field theory},\ }\href
  {https://doi.org/10.1007/s10701-009-9345-2} {\bibfield  {journal} {\bibinfo
  {journal} {Foundations of Physics}\ }\textbf {\bibinfo {volume} {40}},\
  \bibinfo {pages} {746} (\bibinfo {year} {2010})}\BibitemShut {NoStop}%
\bibitem [{\citenamefont {Haldane}(1991)}]{Haldane_exclusion}%
  \BibitemOpen
  \bibfield  {author} {\bibinfo {author} {\bibfnamefont {F.~D.~M.}\
  \bibnamefont {Haldane}},\ }\bibfield  {title} {\bibinfo {title} {``fractional
  statistics'' in arbitrary dimensions: A generalization of the pauli
  principle},\ }\href {https://doi.org/10.1103/PhysRevLett.67.937} {\bibfield
  {journal} {\bibinfo  {journal} {Phys. Rev. Lett.}\ }\textbf {\bibinfo
  {volume} {67}},\ \bibinfo {pages} {937} (\bibinfo {year} {1991})}\BibitemShut
  {NoStop}%
\bibitem [{\citenamefont {Dasni\`eres~de Veigy}\ and\ \citenamefont
  {Ouvry}(1994)}]{PhysRevLett.72.600}%
  \BibitemOpen
  \bibfield  {author} {\bibinfo {author} {\bibfnamefont {A.}~\bibnamefont
  {Dasni\`eres~de Veigy}}\ and\ \bibinfo {author} {\bibfnamefont
  {S.}~\bibnamefont {Ouvry}},\ }\bibfield  {title} {\bibinfo {title} {Equation
  of state of an anyon gas in a strong magnetic field},\ }\href
  {https://doi.org/10.1103/PhysRevLett.72.600} {\bibfield  {journal} {\bibinfo
  {journal} {Phys. Rev. Lett.}\ }\textbf {\bibinfo {volume} {72}},\ \bibinfo
  {pages} {600} (\bibinfo {year} {1994})}\BibitemShut {NoStop}%
\bibitem [{\citenamefont {Ye}\ \emph {et~al.}(2015)\citenamefont {Ye},
  \citenamefont {Marchetti}, \citenamefont {Su},\ and\ \citenamefont
  {Yu}}]{PhysRevB.92.235151}%
  \BibitemOpen
  \bibfield  {author} {\bibinfo {author} {\bibfnamefont {F.}~\bibnamefont
  {Ye}}, \bibinfo {author} {\bibfnamefont {P.~A.}\ \bibnamefont {Marchetti}},
  \bibinfo {author} {\bibfnamefont {Z.~B.}\ \bibnamefont {Su}},\ and\ \bibinfo
  {author} {\bibfnamefont {L.}~\bibnamefont {Yu}},\ }\bibfield  {title}
  {\bibinfo {title} {Hall effect, edge states, and haldane exclusion statistics
  in two-dimensional space},\ }\href
  {https://doi.org/10.1103/PhysRevB.92.235151} {\bibfield  {journal} {\bibinfo
  {journal} {Phys. Rev. B}\ }\textbf {\bibinfo {volume} {92}},\ \bibinfo
  {pages} {235151} (\bibinfo {year} {2015})}\BibitemShut {NoStop}%
\bibitem [{\citenamefont {Murthy}\ and\ \citenamefont
  {Shankar}(1994)}]{murthy1}%
  \BibitemOpen
  \bibfield  {author} {\bibinfo {author} {\bibfnamefont {M.~V.~N.}\
  \bibnamefont {Murthy}}\ and\ \bibinfo {author} {\bibfnamefont
  {R.}~\bibnamefont {Shankar}},\ }\bibfield  {title} {\bibinfo {title} {Haldane
  exclusion statistics and second virial coefficient},\ }\href
  {https://doi.org/10.1103/PhysRevLett.72.3629} {\bibfield  {journal} {\bibinfo
   {journal} {Phys. Rev. Lett.}\ }\textbf {\bibinfo {volume} {72}},\ \bibinfo
  {pages} {3629} (\bibinfo {year} {1994})}\BibitemShut {NoStop}%
\bibitem [{\citenamefont {Wu}\ and\ \citenamefont
  {Yu}(1995)}]{PhysRevLett.75.890}%
  \BibitemOpen
  \bibfield  {author} {\bibinfo {author} {\bibfnamefont {Y.-S.}\ \bibnamefont
  {Wu}}\ and\ \bibinfo {author} {\bibfnamefont {Y.}~\bibnamefont {Yu}},\
  }\bibfield  {title} {\bibinfo {title} {Bosonization of one-dimensional
  exclusons and characterization of luttinger liquids},\ }\href
  {https://doi.org/10.1103/PhysRevLett.75.890} {\bibfield  {journal} {\bibinfo
  {journal} {Phys. Rev. Lett.}\ }\textbf {\bibinfo {volume} {75}},\ \bibinfo
  {pages} {890} (\bibinfo {year} {1995})}\BibitemShut {NoStop}%
\bibitem [{\citenamefont {Wu}\ \emph {et~al.}(2001)\citenamefont {Wu},
  \citenamefont {Yu},\ and\ \citenamefont {Yang}}]{WU2001551}%
  \BibitemOpen
  \bibfield  {author} {\bibinfo {author} {\bibfnamefont {Y.-S.}\ \bibnamefont
  {Wu}}, \bibinfo {author} {\bibfnamefont {Y.}~\bibnamefont {Yu}},\ and\
  \bibinfo {author} {\bibfnamefont {H.-X.}\ \bibnamefont {Yang}},\ }\bibfield
  {title} {\bibinfo {title} {Characterization of one-dimensional luttinger
  liquids in terms of fractional exclusion statistics},\ }\href
  {https://doi.org/https://doi.org/10.1016/S0550-3213(01)00120-1} {\bibfield
  {journal} {\bibinfo  {journal} {Nuclear Physics B}\ }\textbf {\bibinfo
  {volume} {604}},\ \bibinfo {pages} {551} (\bibinfo {year}
  {2001})}\BibitemShut {NoStop}%
\bibitem [{\citenamefont {Rego}\ and\ \citenamefont
  {Kirczenow}(1999)}]{Fractional_thermal}%
  \BibitemOpen
  \bibfield  {author} {\bibinfo {author} {\bibfnamefont {L.~G.~C.}\
  \bibnamefont {Rego}}\ and\ \bibinfo {author} {\bibfnamefont {G.}~\bibnamefont
  {Kirczenow}},\ }\bibfield  {title} {\bibinfo {title} {Fractional exclusion
  statistics and the universal quantum of thermal conductance: A unifying
  approach},\ }\href {https://doi.org/10.1103/PhysRevB.59.13080} {\bibfield
  {journal} {\bibinfo  {journal} {Phys. Rev. B}\ }\textbf {\bibinfo {volume}
  {59}},\ \bibinfo {pages} {13080} (\bibinfo {year} {1999})}\BibitemShut
  {NoStop}%
\bibitem [{\citenamefont {Isakov}\ \emph {et~al.}(1999)\citenamefont {Isakov},
  \citenamefont {Martin},\ and\ \citenamefont {Ouvry}}]{Martin}%
  \BibitemOpen
  \bibfield  {author} {\bibinfo {author} {\bibfnamefont {S.~B.}\ \bibnamefont
  {Isakov}}, \bibinfo {author} {\bibfnamefont {T.}~\bibnamefont {Martin}},\
  and\ \bibinfo {author} {\bibfnamefont {S.}~\bibnamefont {Ouvry}},\ }\bibfield
   {title} {\bibinfo {title} {Conductance and shot noise for particles with
  exclusion statistics},\ }\href {https://doi.org/10.1103/PhysRevLett.83.580}
  {\bibfield  {journal} {\bibinfo  {journal} {Phys. Rev. Lett.}\ }\textbf
  {\bibinfo {volume} {83}},\ \bibinfo {pages} {580} (\bibinfo {year}
  {1999})}\BibitemShut {NoStop}%
\bibitem [{\citenamefont {Wu}(1994)}]{Wu}%
  \BibitemOpen
  \bibfield  {author} {\bibinfo {author} {\bibfnamefont {Y.-S.}\ \bibnamefont
  {Wu}},\ }\bibfield  {title} {\bibinfo {title} {Statistical distribution for
  generalized ideal gas of fractional-statistics particles},\ }\href
  {https://doi.org/10.1103/PhysRevLett.73.922} {\bibfield  {journal} {\bibinfo
  {journal} {Phys. Rev. Lett.}\ }\textbf {\bibinfo {volume} {73}},\ \bibinfo
  {pages} {922} (\bibinfo {year} {1994})}\BibitemShut {NoStop}%
\bibitem [{\citenamefont {Nayak}\ and\ \citenamefont {Wilczek}(1994)}]{Chetan}%
  \BibitemOpen
  \bibfield  {author} {\bibinfo {author} {\bibfnamefont {C.}~\bibnamefont
  {Nayak}}\ and\ \bibinfo {author} {\bibfnamefont {F.}~\bibnamefont
  {Wilczek}},\ }\bibfield  {title} {\bibinfo {title} {Exclusion statistics:
  Low-temperature properties, fluctuations, duality, and applications},\ }\href
  {https://doi.org/10.1103/PhysRevLett.73.2740} {\bibfield  {journal} {\bibinfo
   {journal} {Phys. Rev. Lett.}\ }\textbf {\bibinfo {volume} {73}},\ \bibinfo
  {pages} {2740} (\bibinfo {year} {1994})}\BibitemShut {NoStop}%
\bibitem [{\citenamefont {ISAKOV}(1994)}]{ISAKOV1}%
  \BibitemOpen
  \bibfield  {author} {\bibinfo {author} {\bibfnamefont {S.~B.}\ \bibnamefont
  {ISAKOV}},\ }\bibfield  {title} {\bibinfo {title} {Generalization of
  statistics for several species of identical particles},\ }\href
  {https://doi.org/10.1142/S0217984994000327} {\bibfield  {journal} {\bibinfo
  {journal} {Modern Physics Letters B}\ }\textbf {\bibinfo {volume} {08}},\
  \bibinfo {pages} {319} (\bibinfo {year} {1994})},\ \Eprint
  {https://arxiv.org/abs/https://doi.org/10.1142/S0217984994000327}
  {https://doi.org/10.1142/S0217984994000327} \BibitemShut {NoStop}%
\bibitem [{\citenamefont {Viola}\ \emph {et~al.}(2012)\citenamefont {Viola},
  \citenamefont {Das}, \citenamefont {Grosfeld},\ and\ \citenamefont
  {Stern}}]{Viola}%
  \BibitemOpen
  \bibfield  {author} {\bibinfo {author} {\bibfnamefont {G.}~\bibnamefont
  {Viola}}, \bibinfo {author} {\bibfnamefont {S.}~\bibnamefont {Das}}, \bibinfo
  {author} {\bibfnamefont {E.}~\bibnamefont {Grosfeld}},\ and\ \bibinfo
  {author} {\bibfnamefont {A.}~\bibnamefont {Stern}},\ }\bibfield  {title}
  {\bibinfo {title} {Thermoelectric probe for neutral edge modes in the
  fractional quantum hall regime},\ }\href
  {https://doi.org/10.1103/PhysRevLett.109.146801} {\bibfield  {journal}
  {\bibinfo  {journal} {Phys. Rev. Lett.}\ }\textbf {\bibinfo {volume} {109}},\
  \bibinfo {pages} {146801} (\bibinfo {year} {2012})}\BibitemShut {NoStop}%
\bibitem [{\citenamefont {Gurman}\ \emph {et~al.}(2012)\citenamefont {Gurman},
  \citenamefont {Sabo}, \citenamefont {Heiblum}, \citenamefont {Umansky},\ and\
  \citenamefont {Mahalu}}]{Gurman2012}%
  \BibitemOpen
  \bibfield  {author} {\bibinfo {author} {\bibfnamefont {I.}~\bibnamefont
  {Gurman}}, \bibinfo {author} {\bibfnamefont {R.}~\bibnamefont {Sabo}},
  \bibinfo {author} {\bibfnamefont {M.}~\bibnamefont {Heiblum}}, \bibinfo
  {author} {\bibfnamefont {V.}~\bibnamefont {Umansky}},\ and\ \bibinfo {author}
  {\bibfnamefont {D.}~\bibnamefont {Mahalu}},\ }\bibfield  {title} {\bibinfo
  {title} {Extracting net current from an upstream neutral mode in the
  fractional quantum hall regime},\ }\href {https://doi.org/10.1038/ncomms2305}
  {\bibfield  {journal} {\bibinfo  {journal} {Nature Communications}\ }\textbf
  {\bibinfo {volume} {3}},\ \bibinfo {pages} {1289} (\bibinfo {year}
  {2012})}\BibitemShut {NoStop}%
\bibitem [{\citenamefont {Krive}\ and\ \citenamefont {Mucciolo}(1999)}]{Krive}%
  \BibitemOpen
  \bibfield  {author} {\bibinfo {author} {\bibfnamefont {I.~V.}\ \bibnamefont
  {Krive}}\ and\ \bibinfo {author} {\bibfnamefont {E.~R.}\ \bibnamefont
  {Mucciolo}},\ }\bibfield  {title} {\bibinfo {title} {Transport properties of
  quasiparticles with fractional exclusion statistics},\ }\href
  {https://doi.org/10.1103/PhysRevB.60.1429} {\bibfield  {journal} {\bibinfo
  {journal} {Phys. Rev. B}\ }\textbf {\bibinfo {volume} {60}},\ \bibinfo
  {pages} {1429} (\bibinfo {year} {1999})}\BibitemShut {NoStop}%
\bibitem [{\citenamefont {Paulsson}\ and\ \citenamefont
  {Datta}(2003)}]{Supriyo}%
  \BibitemOpen
  \bibfield  {author} {\bibinfo {author} {\bibfnamefont {M.}~\bibnamefont
  {Paulsson}}\ and\ \bibinfo {author} {\bibfnamefont {S.}~\bibnamefont
  {Datta}},\ }\bibfield  {title} {\bibinfo {title} {Thermoelectric effect in
  molecular electronics},\ }\href {https://doi.org/10.1103/PhysRevB.67.241403}
  {\bibfield  {journal} {\bibinfo  {journal} {Phys. Rev. B}\ }\textbf {\bibinfo
  {volume} {67}},\ \bibinfo {pages} {241403} (\bibinfo {year}
  {2003})}\BibitemShut {NoStop}%
\bibitem [{\citenamefont {Alhassid}(2000)}]{Alhassid}%
  \BibitemOpen
  \bibfield  {author} {\bibinfo {author} {\bibfnamefont {Y.}~\bibnamefont
  {Alhassid}},\ }\bibfield  {title} {\bibinfo {title} {The statistical theory
  of quantum dots},\ }\href {https://doi.org/10.1103/RevModPhys.72.895}
  {\bibfield  {journal} {\bibinfo  {journal} {Rev. Mod. Phys.}\ }\textbf
  {\bibinfo {volume} {72}},\ \bibinfo {pages} {895} (\bibinfo {year}
  {2000})}\BibitemShut {NoStop}%
\bibitem [{\citenamefont {Nakpathomkun}\ \emph {et~al.}(2010)\citenamefont
  {Nakpathomkun}, \citenamefont {Xu},\ and\ \citenamefont
  {Linke}}]{Nakpathomkun}%
  \BibitemOpen
  \bibfield  {author} {\bibinfo {author} {\bibfnamefont {N.}~\bibnamefont
  {Nakpathomkun}}, \bibinfo {author} {\bibfnamefont {H.~Q.}\ \bibnamefont
  {Xu}},\ and\ \bibinfo {author} {\bibfnamefont {H.}~\bibnamefont {Linke}},\
  }\bibfield  {title} {\bibinfo {title} {Thermoelectric efficiency at maximum
  power in low-dimensional systems},\ }\href
  {https://doi.org/10.1103/PhysRevB.82.235428} {\bibfield  {journal} {\bibinfo
  {journal} {Phys. Rev. B}\ }\textbf {\bibinfo {volume} {82}},\ \bibinfo
  {pages} {235428} (\bibinfo {year} {2010})}\BibitemShut {NoStop}%
\bibitem [{\citenamefont {Kennes}\ \emph {et~al.}(2013)\citenamefont {Kennes},
  \citenamefont {Schuricht},\ and\ \citenamefont {Meden}}]{Kennes_2013}%
  \BibitemOpen
  \bibfield  {author} {\bibinfo {author} {\bibfnamefont {D.~M.}\ \bibnamefont
  {Kennes}}, \bibinfo {author} {\bibfnamefont {D.}~\bibnamefont {Schuricht}},\
  and\ \bibinfo {author} {\bibfnamefont {V.}~\bibnamefont {Meden}},\ }\bibfield
   {title} {\bibinfo {title} {Efficiency and power of a thermoelectric quantum
  dot device},\ }\href {https://doi.org/10.1209/0295-5075/102/57003} {\bibfield
   {journal} {\bibinfo  {journal} {Europhysics Letters}\ }\textbf {\bibinfo
  {volume} {102}},\ \bibinfo {pages} {57003} (\bibinfo {year}
  {2013})}\BibitemShut {NoStop}%
\bibitem [{\citenamefont {Benenti}\ \emph {et~al.}(2017)\citenamefont
  {Benenti}, \citenamefont {Casati}, \citenamefont {Saito},\ and\ \citenamefont
  {Whitney}}]{Whitney_review}%
  \BibitemOpen
  \bibfield  {author} {\bibinfo {author} {\bibfnamefont {G.}~\bibnamefont
  {Benenti}}, \bibinfo {author} {\bibfnamefont {G.}~\bibnamefont {Casati}},
  \bibinfo {author} {\bibfnamefont {K.}~\bibnamefont {Saito}},\ and\ \bibinfo
  {author} {\bibfnamefont {R.}~\bibnamefont {Whitney}},\ }\bibfield  {title}
  {\bibinfo {title} {Fundamental aspects of steady-state conversion of heat to
  work at the nanoscale},\ }\href
  {https://doi.org/https://doi.org/10.1016/j.physrep.2017.05.008} {\bibfield
  {journal} {\bibinfo  {journal} {Physics Reports}\ }\textbf {\bibinfo {volume}
  {694}},\ \bibinfo {pages} {1} (\bibinfo {year} {2017})},\ \bibinfo {note}
  {fundamental aspects of steady-state conversion of heat to work at the
  nanoscale}\BibitemShut {NoStop}%
\bibitem [{\citenamefont {Whitney}(2014)}]{boxcar}%
  \BibitemOpen
  \bibfield  {author} {\bibinfo {author} {\bibfnamefont {R.~S.}\ \bibnamefont
  {Whitney}},\ }\bibfield  {title} {\bibinfo {title} {Most efficient quantum
  thermoelectric at finite power output},\ }\href
  {https://doi.org/10.1103/PhysRevLett.112.130601} {\bibfield  {journal}
  {\bibinfo  {journal} {Phys. Rev. Lett.}\ }\textbf {\bibinfo {volume} {112}},\
  \bibinfo {pages} {130601} (\bibinfo {year} {2014})}\BibitemShut {NoStop}%
\bibitem [{\citenamefont {Kane}\ and\ \citenamefont
  {Fisher}(1996)}]{kane1996thermal}%
  \BibitemOpen
  \bibfield  {author} {\bibinfo {author} {\bibfnamefont {C.~L.}\ \bibnamefont
  {Kane}}\ and\ \bibinfo {author} {\bibfnamefont {M.~P.~A.}\ \bibnamefont
  {Fisher}},\ }\bibfield  {title} {\bibinfo {title} {Thermal transport in a
  luttinger liquid},\ }\href {https://doi.org/10.1103/PhysRevLett.76.3192}
  {\bibfield  {journal} {\bibinfo  {journal} {Phys. Rev. Lett.}\ }\textbf
  {\bibinfo {volume} {76}},\ \bibinfo {pages} {3192} (\bibinfo {year}
  {1996})}\BibitemShut {NoStop}%
\bibitem [{\citenamefont {Grosfeld}\ and\ \citenamefont
  {Das}(2009)}]{Etyan_Sourin_2009}%
  \BibitemOpen
  \bibfield  {author} {\bibinfo {author} {\bibfnamefont {E.}~\bibnamefont
  {Grosfeld}}\ and\ \bibinfo {author} {\bibfnamefont {S.}~\bibnamefont {Das}},\
  }\bibfield  {title} {\bibinfo {title} {Probing the neutral edge modes in
  transport across a point contact via thermal effects in the read-rezayi
  non-abelian quantum hall states},\ }\href
  {https://doi.org/10.1103/PhysRevLett.102.106403} {\bibfield  {journal}
  {\bibinfo  {journal} {Phys. Rev. Lett.}\ }\textbf {\bibinfo {volume} {102}},\
  \bibinfo {pages} {106403} (\bibinfo {year} {2009})}\BibitemShut {NoStop}%
\bibitem [{\citenamefont {Hajiloo}\ \emph {et~al.}(2020)\citenamefont
  {Hajiloo}, \citenamefont {S\'anchez}, \citenamefont {Whitney},\ and\
  \citenamefont {Splettstoesser}}]{Whitney}%
  \BibitemOpen
  \bibfield  {author} {\bibinfo {author} {\bibfnamefont {F.}~\bibnamefont
  {Hajiloo}}, \bibinfo {author} {\bibfnamefont {R.}~\bibnamefont {S\'anchez}},
  \bibinfo {author} {\bibfnamefont {R.~S.}\ \bibnamefont {Whitney}},\ and\
  \bibinfo {author} {\bibfnamefont {J.}~\bibnamefont {Splettstoesser}},\
  }\bibfield  {title} {\bibinfo {title} {Quantifying nonequilibrium
  thermodynamic operations in a multiterminal mesoscopic system},\ }\href
  {https://doi.org/10.1103/PhysRevB.102.155405} {\bibfield  {journal} {\bibinfo
   {journal} {Phys. Rev. B}\ }\textbf {\bibinfo {volume} {102}},\ \bibinfo
  {pages} {155405} (\bibinfo {year} {2020})}\BibitemShut {NoStop}%
\bibitem [{\citenamefont {Isakov}(1994)}]{Isakov}%
  \BibitemOpen
  \bibfield  {author} {\bibinfo {author} {\bibfnamefont {S.~B.}\ \bibnamefont
  {Isakov}},\ }\bibfield  {title} {\bibinfo {title} {Statistical mechanics for
  a class of quantum statistics},\ }\href
  {https://doi.org/10.1103/PhysRevLett.73.2150} {\bibfield  {journal} {\bibinfo
   {journal} {Phys. Rev. Lett.}\ }\textbf {\bibinfo {volume} {73}},\ \bibinfo
  {pages} {2150} (\bibinfo {year} {1994})}\BibitemShut {NoStop}%
\bibitem [{\citenamefont {Sivan}\ and\ \citenamefont {Imry}(1986)}]{Imry}%
  \BibitemOpen
  \bibfield  {author} {\bibinfo {author} {\bibfnamefont {U.}~\bibnamefont
  {Sivan}}\ and\ \bibinfo {author} {\bibfnamefont {Y.}~\bibnamefont {Imry}},\
  }\bibfield  {title} {\bibinfo {title} {Multichannel landauer formula for
  thermoelectric transport with application to thermopower near the mobility
  edge},\ }\href {https://doi.org/10.1103/PhysRevB.33.551} {\bibfield
  {journal} {\bibinfo  {journal} {Phys. Rev. B}\ }\textbf {\bibinfo {volume}
  {33}},\ \bibinfo {pages} {551} (\bibinfo {year} {1986})}\BibitemShut
  {NoStop}%
\end{thebibliography}%
\widetext{
\appendix
\section{Useful integrals}
The occupation function in eq.~(1) of the main text has the form ~\cite{Wu,Chetan,Isakov}:
\begin{align}
    \eta_g(E;\mu,T)=\frac{1}{\mathcal{W}(x,g)+g},
    \label{Eq:distA}
\end{align}
 and $\mathcal{W}(x,g)$ satisfies the functional equation
\begin{align}
    \mathcal{W}(x,g)^g\,[1+\mathcal{W}(x,g)]^{1-g}=e^{x}.
    \label{Eq:WdefineA}
\end{align}
where $g$ is the statistical parameter, $x=\beta(E-\mu)$, $\beta=1/(k_\text{B} T)$, $E$ denotes energy, $T$ is temperature, $\mu$ is the chemical potential. By taking logarithms on both sides
\begin{align}
   x(\mathcal{W},g)= \beta(E-\mu)=g\ln \mathcal{W}+(1-g)\ln(1+ \mathcal{W}).
\end{align}
By varying $E$, we can write 
\begin{align}
   dE=\frac{1}{\beta}\frac{\mathcal{W}+g}{\mathcal{W}(\mathcal{W}+1)}d\mathcal{W}.
\end{align}
By differentiating eq.~(\ref{Eq:distA}) with respect to $E$, we obtain:
\begin{align}
   \eta_g'(E;\mu,T)=- \beta\frac{\mathcal{W}(\mathcal{W}+1)}{(\mathcal{W}+g)^3}.
\end{align}
Assuming $E(0)=-\infty$, which represents the lowest energy, valid as long as it remains unaffected by the excitation energy window, we get~\cite{Fractional_thermal}
\begin{align}
   &(a)    \int_{E(0)}^\infty dE\big(-\eta_g'(E;\mu,T)\big)=\int_0^\infty\frac{d\mathcal{W}}{(\mathcal{W}+g)^2}=\frac{1}{g},\label{eq:a}\\
    &(b)    \int_{E(0)}^\infty dE(E-\mu) \big(-\eta_g'(E;\mu,T)\big)=\frac{1}{\beta}\int_0^\infty\frac{g\ln \mathcal{W}+(1-g)\ln(1+ \mathcal{W})}{(\mathcal{W}+g)^2}d\mathcal{W}=0,\label{eq:b}\\
     &(c)    \int_{E(0)}^\infty dE(E-\mu)^2 \big(-\eta_g'(E;\mu,T)\big)=\frac{1}{\beta^2}\int_0^\infty\frac{(g\ln \mathcal{W}+(1-g)\ln(1+ \mathcal{W}))^2}{(\mathcal{W}+g)^2}d\mathcal{W}=\frac{\pi^2}{3\beta^2}.\label{eq:c}
\end{align}
\section{Linear transport coefficients}\label{sec:Linear_transport_coefficients}
We have considered a QPC setup between two counterpropagating one-dimensional channels, characterized by particles of statistical parameter $g$. The two incoming edges are maintained at chemical potentials (temperatures), $\mu_1 (T_1)$ and $\mu_2 (T_2)$, respectively. The chemical potentials are related to the voltages $V_1$ and $V_2$ via $\mu_i=qV_i$, where $q$ denotes the charge of the particles. We assume $\mu_{i}=\mu+\delta \mu_i$ and $T_{i}=T+\delta T_i$ for $i=1,2$, where $\mu$($T$) denotes the average chemical potential (temperature) of the system and $\delta\mu_i, k_\text{B}\,\delta T_i$ are assumed to be very small compared to $k_\text{B}\, T$ and $\mu$. We consider the linear response limit, where the applied chemical potential bias $\mu_{1}-\mu_{2}=\delta \mu_{1}-\delta\mu_{2}=\Delta \mu=q\Delta V  \longrightarrow 0$ and temperature difference $T_1-T_2=\delta T_1-\delta T_2=\Delta T\longrightarrow 0$. The QPC is considered to be partially transmitting, where particles incident along the incoming edges are transmitted or reflected to the outgoing edge with energy-dependent probabilities $t(E) $ and $r(E)$, respectively, satisfying $t(E)+r(E) =1$ to ensure current conservation.

Since $\eta_g(E;\mu,T)$ exhibits dependence in the form $\eta_g\big((E-\mu)/T\big)$, the first-order expansion of the occupation function for the $i^{th}$ channel, $\eta_g(E;\mu_i,T_i)$, in $\delta\mu_i$ and $\delta T_i$ can be written to be equal to
\begin{align}
    \eta_g(E;\mu,T) +\Big(\delta\mu_i+\frac{E-\mu}{T}\,\delta T_i\Big)\,\big(-\eta_g'(E;\mu,T)\big)\label{Eq:Dist_expansion},
\end{align}
where $\eta_g'(E;\mu,T)$ represents the first order derivative of $\eta_g(E;\mu,T)$ with respect to $E$.

It has been demonstrated in Ref.~[\onlinecite{Krive}] that the transport coefficients for particles obeying exclusion statistics follow the same general expressions as those for particles with conventional statistics. The net charge current going from left to right of the system is 
\begin{align}
I_{net} =q\int_{E(0)}^{\infty}\frac{dE}{2\pi}\mathcal{D}\,v \big(\eta_g(E;\mu_1,T_1) - \big[r(E)\eta_g(E;\mu_1,T_1)+t(E)\eta_g(E;\mu_2,T_2)\big]\big).
\end{align}
Here the velocity of particles is $v=\hbar^{-1}\partial E/\partial k$ and the density of states ($\mathcal{D}=\partial k/\partial E$) are the same on both edges and assumed to be energy independent. Using  eq.~(\ref{Eq:Dist_expansion}), this simplifies to
\begin{align}
I_{net} =\frac{q}{h}\bigg[\Delta \mu\int_{E(0)}^\infty dE \,t(E)\big(-\eta_g'(E;\mu,T)\big)+\frac{\Delta T}{T}\int_{E(0)}^\infty dE \,t(E)(E-\mu)\big(-\eta_g'(E;\mu,T)\big)\bigg].
\label{Eq:Net_I_FESA}
\end{align}
To calculate the heat current, we begin by considering the right and left moving entropy currents~\cite{Imry}, given as
  \begin{align}
  & J_{\mathcal{S}i}=-k_{\text{B}}\int_{E(0)}^\infty\frac{dE}{2\pi}\mathcal{D}\,v\Big(\eta_g(E;\mu_1,T_1)\ln\eta_g(E;\mu_1,T_1)   +\big[1-g\,\eta_g(E;\mu_1,T_1)\big]\ln\big[1-g\,\eta_g(E;\mu_1,T_1)\big]\nonumber\\
   &\hspace{7cm}-\big[1+(1-g)\eta_g(E;\mu_1,T_1)\big]\ln\big[1+(1-g)\eta_g(E;\mu_1,T_1)\big]\Big),\\
    &J_{\mathcal{S}o}=-k_{\text{B}}\int_{E(0)}^\infty\frac{dE}{2\pi}\mathcal{D}\,v\Big( \big[r(E)\eta_g(E;\mu_1,T_1)+t(E)\eta_g(E;\mu_2,T_2)\big]\ln\big[r(E)\eta_g(E;\mu_1,T_1)+t(E)\eta_g(E;\mu_2,T_2)\big]\nonumber\\
    &\hspace{1cm}+\big[1-g\big(r(E)\eta_g(E;\mu_1,T_1)+t(E)\eta_g(E;\mu_2,T_2)\big)\big]\ln\big[1-g\big(r(E)\eta_g(E;\mu_1,T_1)+t(E)\eta_g(E;\mu_2,T_2)\big)\big] \nonumber\\
 &\hspace{1cm}-\big[1+(1-g)\big(r(E)\eta_g(E;\mu_1,T_1)+t(E)\eta_g(E;\mu_2,T_2)\big)\big]\ln\big[1+(1-g)\big(r(E)\eta_g(E;\mu_1,T_1)+t(E)\eta_g(E;\mu_2,T_2)\big)\big]\Big).
\end{align}
Hence, the net entropy current flowing from left to write is $J_{\mathcal{S},net}=J_{\mathcal{S}i}-J_{\mathcal{S}o}$. In linear response and using eq.~(\ref{Eq:distA}),(\ref{Eq:WdefineA}), it can be shown to be equal to 
\begin{align}
    J_{\mathcal{S},net}=k_{\text{B}}\int_{E(0)}^\infty\frac{dE}{2\pi}\mathcal{D}\,v\,\beta(E-\mu_1)\,t(E)\big(\eta_g(E;\mu_1,T_1)- \eta_g(E;\mu_2,T_2)\big)
\end{align}
Thus obtain the expression for the net heat current as
\begin{align}
J_{net} =T J_{\mathcal{S},net}=\int_{E(0)}^{\infty}\frac{dE}{2\pi} (E-\mu_1)\mathcal{D}\,v \big(\eta_g(E;\mu_1,T_1)- \eta_g(E;\mu_2,T_2)\big).
\end{align}
In the linear response limit, this simplifies to 
\begin{align}
J_{net} =&\frac{1}{h}\bigg[\Delta \mu\int_{E(0)}^\infty dE \,t(E)(E-\mu)\big(-\eta_g'(E;\mu,T)\big)+\frac{\Delta T}{T}\int_{E(0)}^\infty dE \,t(E)(E-\mu)^2\big(-\eta_g'(E;\mu,T)\big)\bigg].\label{Eq:Net_J_FESA}
\end{align}
Since we are considering the linear response limit, which is first order in the chemical potential bias and temperature difference, this framework does not account for power generation or absorption in the system, such as Joule heating, as these effects scale quadratically with these parameters and lie beyond the scope of this approximation. Eq.~(\ref{Eq:Net_I_FESA}) and (\ref{Eq:Net_J_FESA}) together can be rewritten in a  matrix form 
\begin{align}
    \begin{pmatrix}
        I_{net}\\J_{net}\end{pmatrix}=\begin{pmatrix}          
       q^2 \mathcal{L}_0 &q \mathcal{L}_1\\
        q\mathcal{L}_1 & \mathcal{L}_2
    \end{pmatrix}\begin{pmatrix}
        \Delta\mu/q\\ \Delta T /T\end{pmatrix},\label{Eq:Onsager_matrixA}
\end{align}
where
\begin{align}
    \mathcal{L}_\alpha=\frac{1}{h}\int_{E(0)}^\infty dE \,t(E)(E-\mu)^\alpha\big(-\eta_g'(E;\mu,T)\big),\hspace{2cm}\alpha=0,1,2.
\end{align}
Eq.~(\ref{Eq:Onsager_matrixA}) explicitly confirms that the Onsager relation is satisfied, ensuring the reciprocity of the transport coefficients in accordance with fundamental thermodynamic principles. Generally, in experiments, $I_{net}$ and $\Delta T$ serve as independent parameters, and correspondingly eq.~(\ref{Eq:Onsager_matrixA}) becomes
\begin{align}
    \begin{pmatrix}
        \Delta V\\J_{net}\end{pmatrix}=\begin{pmatrix}          
      1/G & -S\\
        \Pi & \kappa
    \end{pmatrix}\begin{pmatrix}
        I_{net}\\ \Delta T \end{pmatrix},\label{Eq:Onsager_matrix}
\end{align}
where the definitions of the transport coefficients are:
\begin{align}
    &\text{Longitudinal charge conductance }(G)=\frac{I_{net}}{\Delta V}\Big|_{\Delta T=0}=q^2\mathcal{L}_0;\\
       &\text{Longitudinal thermal conductance }(\kappa)=\frac{J_{net}}{\Delta T}\Big|_{I_{net}=0}=\frac{1}{T}\Big(\mathcal{L}_2-\frac{\mathcal{L}_1^2}{\mathcal{L}_0}\Big);\\
        &\text{Seebeck coefficient }(S) =-\frac{\Delta V}{\Delta T}\Big|_{I_{net}=0}=\frac{1}{qT}\frac{\mathcal{L}_1}{\mathcal{L}_0};\\
     &\text{Peltier coefficient }(\Pi) =\frac{J_{net}}{I_{net}}\Big|_{\Delta T=0}=\frac{1}{q}\frac{\mathcal{L}_1}{\mathcal{L}_0}.
\end{align}
This clearly shows the Onsager relation through $S=\Pi/T$, where the Seebeck coefficient is identified as the entropy transported per unit charge through the system.\\
The expression for Lorenz number is given by
\begin{align}
    L=\frac{\kappa}{G T}=\frac{1}{q^2T^2}\Big(\frac{\mathcal{L}_2}{\mathcal{L}_0}-\frac{\mathcal{L}_1^2}{\mathcal{L}_0^2}\Big),\label{eq:LorenzA}
\end{align}
\end{document}